# Influence of resonances on the $^{11}$B(n,γ)$^{12}$B capture reaction rate. II. Capture to the first excited state of $^{12}$B.

Dubovichenko S.B.[1,2,*], Burkova N.A.[2], Dzhazairov-Kakhramanov A.V.[1,*]

[1]Fesenkov Astrophysical Institute "NCSRT" ASA MDASI RK, 050020, Almaty, Kazakhstan
[2]al-Farabi Kazakh National University, 050040, Almaty, Kazakhstan

**Abstract:** Within the framework of the modified potential cluster model with a classification of orbital states according to Young diagrams, the possibility of prediction absentee experimental data for the total cross sections of the radiative neutron capture on $^{11}$B to the first excited state $^{12}$B at 0.95 MeV ($2^+$) for reaction energy of 10 meV (1 meV = $10^{-3}$ eV) to 7 MeV. The reaction rate in the temperature range of 0.01 to 10.0 $T_9$ is calculated on the basis of obtained cross sections, which take into account resonances up to 5 MeV. It is shown that low-lying resonances exercise a significant influence to the capture reaction rate. The approximation of the calculation reaction rate is carried out by the simple analytic formula.

## 1. Introduction

Earlier, the neutron capture process on $^{11}$B to the ground states (GS) of $^{12}$B was considered by us within the framework of the modified potential cluster model (MPCM) in [1]. It is possible, on the basis of the MPCM, to correctly describe available experimental data for capture to the GS at energies of 25.3 meV to 60 keV. Continuing studying of the radiative capture process in the framework of the MPCM [2,3], let us consider $n^{11}\text{B} \to \gamma^{12}\text{B}$ capture reaction to the first excited state (FES) of $^{12}$B at $J^\pi = 2^+$ with excitation energy of 0.95 MeV and carry out the account of resonance states (RS) in the continuous spectrum up to 5 MeV. In the present paper, evidently, the capture to the FES of $^{12}$B taken into account resonances in the scattering process of initial particles in the input channel is considered first. Our calculations, intrinsically, are predictive and estimative for argumentation and organization of new experiments for the neutron capture on $^{11}$B. Such results can also initiate essential overestimation of the rate of synthesis boron isotopes. It allows one to estimate how the balance of elements can change at synthesis of these nuclei in future.

  In the present calculations we will use new data on spectra of $^{12}$B from work [4], comparably with our previous papers [5], where results of earlier review [6] were used. Besides, in works [5] old data on asymptotic constants (AC) of the $n^{11}$B system were used, which did not allow to describe available experimental data. The considered here reaction $n^{11}\text{B} \to \gamma^{12}\text{B}$ was studied earlier in [7], where for the construction of the $n^{11}$B interactions potentials the thermal neutron capture cross section $\sigma_{th}$ equals 5.5 mb was used [8]. At the same time, more modern data from [9] were used in our work [1], where the value $\sigma_{th}$ = 9.09(10) mb is given.

---

[*] Corresponding authors: albert-j@yandex.kz, dubovichenko@gmail.com

## 2. Calculation methods and state structure

Total cross sections for the radiative capture σ($NJ,J_f$) for $EJ$ and $MJ$ transitions in the potential cluster model are given, for example, in [10] or [1–3]. The values from data base [11] and work [12] were used for cluster magnetic moments, that is, $μ_n$ = -1.91304272$μ_0$ and $μ(^{11}B)$ = 2.6887$μ_0$. The next mass values of particles were used in calculations: $m_n$ = 1.00866491597 amu [11], $m(^{11}B)$ = 11.0093052 amu [13], and constant $ℏ^2/m_0$ is equal to 41.4686 MeV·fm$^2$, where $m_0$ is atomic mass unit (amu). Calculation methods in the frame of the MPCM of other values, for example, root-mean-square mass and charge radii or binding energy, which are considered further, are given, for example, in [14].

The used intercluster potentials for the considered 35 cluster nuclear systems do not contain ambiguities and, as it was shown in [2,3] and [15] (see also refs. in it), allow us to correctly describe the total cross sections of many processes of radiative capture. The potentials of the bound states (BS), namely FES, should correctly describe the known values of the AC, which is associated with the asymptotic normalization coefficient (ANC) of the $A_{NC}$ usually extracted from the experiment as follows [16]

$$A_{NC}^2 = S_f \cdot C^2 \qquad (1)$$

where $S_f$ is the spectroscopic factor of the channel and $C$ is the dimensional AC, expressed in fm$^{-1/2}$ and determined from the relation

$$χ_L(r) = C \cdot W_{-ηL+1/2}(2k_0 r). \qquad (2)$$

$C$ is related to the dimensionless AC $C_w$ [17], used by us, as follows: $C = \sqrt{2k_0} C_w$, and the dimensionless constant $C_w$ can be defined by the expression [17]

$$χ_L(r) = \sqrt{2k_0} \cdot C_w \cdot W_{-ηL+1/2}(2k_0 r), \qquad (3)$$

where $χ_L(r)$ is the numerical wave function of the bound state obtained from the solution of the radial Schrödinger equation and normalized to unit, $W_{-ηL+1/2}$ is the Whittaker function of the bound state, which determines the asymptotic behavior of the wave function (WF) and is a solution of the same equation without nuclear potential, that is, at large distances $r = R$, $k_0$ is the wave number caused by the channel binding energy, η is the Coulomb parameter, which is equal to zero in this case, and $L$ is the orbital momentum of this BS.

Let us assume furthermore that for $^{11}$B (spin and isospin of $^{11}$B are equal to $J^π,T$ = 3/2$^-$,1/2 [4]) it is possible to take Young diagram in the form {443}, therefore for the $n^{11}$B system we have {1} × {443} → {543} + {444} + {4431} [18]. First of the obtained diagrams compatible with orbital momenta $L$ = 1,2,3,4 and is forbidden (FS), because it cannot be five nucleons in the $s$-shell [19], the second diagram {444} apparently corresponds to the allowed state (AS) compatible with orbital momenta $L$ = 0,2,4, and the third {4431}, also allowed, compatible with $L$ = 1,2,3 [1,19].

Thus, limiting only by the low partial waves with orbital momenta $L$ = 0,1,2,3, it is possible to say that there is only AS for diagram {444} in the $S$ wave potential of the $n^{11}$B system. There are forbidden {543} and allowed {4431} states in $P$ waves. In particular, the GS of $^{12}$B with momentum $J^\pi = 1^+$ corresponds to the $P_1$ wave with {4431}, lying at binding energy of the $n^{11}$B system of -3.370 MeV [4]. FS with the diagram {543} and AS at {4431}+{444} correspond to the $D$ waves. FS with {543} and AS with {4431} correspond to the $F$ waves. These AS for scattering potentials can lie in the continuous spectrum and be unbounded. The AS can correspond to the FES of $^{12}$B with momenta $J^\pi = 2^+$, lying at the binding energy of the $n^{11}$B system of -2.4169 MeV [4].

Besides, some $n^{11}$B scattering states and BSs can be mixed by spin channel with $S = 1$ and 2. Therefore, both of spin states $^3P_2$ and $^5P_2$ can give a contribution in the WF of the FES, and the FES should be considered as the $^{3+5}P_2$ mixture. The same is related to the GS of $^{12}$B in the $n^{11}$B channel, which is the $^{3+5}P_1$ mixture [1]. The present model does not allow to extract in the WF states with $S = 1$ and 2, therefore the GS function is the $^{3+5}P_1$ level and the FES is the $^{3+5}P_2$ level and are obtained from the solution of the Schrödinger equations with the given potential. The similar situation was, for example, at the $^{15}$N$(n,\gamma)^{16}$N and $^7$Li$(n,\gamma)^8$Li captures, when the GS WF was presented as the mixture of $^{3+5}P_2$ waves [2,3].

## 3. Possible transitions in the $n^{11}$B system to the FES of $^{12}$B

Because, here the FES of $^{12}$B is compared to the $^{3+5}P_2$ level, and it is possible to consider $E1$ transitions from nonresonance, as we think, at low energies of the $^3S_1$ and $^5S_2$ waves of the $n^{11}$B scattering to different components of WF of the FES of $^{12}$B in the $n^{11}$B channel $^3S_1 \xrightarrow{E1} {}^3P_2$; $^5S_2 \xrightarrow{E1} {}^5P_2$. Cross sections of such process can be written as a sum $\sigma(E1) = \sigma(^3S_1 \to {}^3P_2) + \sigma(^5S_2 \to {}^5P_2)$, therefore transitions occur from different partial scattering wave to the same FES of $^{12}$B, which is compared to the mixed by spin WF $^{3+5}P_2$.

Let us consider now available excited states in $^{12}$B nucleus, but bound in the $n^{11}$B channel.

1. There is first excited state (1$^{st}$ ES), but not bound in this channel, with momentum of $J^\pi = 2^+$, which can be matched to the $^{3+5}P_2$ wave with the bound FS, at the excitation energy of 0.95314(60) MeV or -2.41686(60) MeV [4] relatively to the threshold of the $n^{11}$B channel.

2. Second excited state (2$^{nd}$ ES) at excitation energy of 1.67365(60) MeV [6] relatively to the GS or -1.69635(60) MeV relatively to the threshold of the $n^{11}$B channel has $J^\pi = 2^-$, and it can be matched to the $^5S_2$ wave without FS. In this case, also $^{3+5}D_2$ wave with the FS is possible.

3. Third excited state (3$^{rd}$ ES) at excitation energy of 2.6208(12) MeV [4] or -0.7492(12) MeV relatively to the threshold of the $n^{11}$B channel has $J^\pi = 1^-$, and it can be matched to the triplet $^3S_1$ wave without forbidden BS. In this case, also $^{3+5}D_1$ wave with the FS is possible.

4. Fourth excited state (4$^{th}$ ES) at excitation energy of 2.723(11) MeV [4] or -0.647(11) MeV [4] relatively to the threshold of the $n^{11}$B channel has $J^\pi = 0^+$, and it can be matched to the triplet $^3P_0$ wave with the bound FS.

The spectrum of ES considered above is shown in Fig. 1a [4].

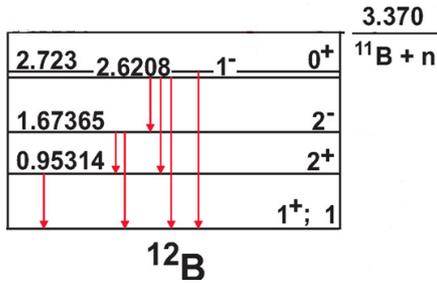

Fig. 1a. Spectrum of excited states of $^{12}$B [4].

Furthermore, the potentials of these ESs were constructed, which are using for description of the scattering processes in these partial waves that do not include resonances [1]. However, in the $^{3+5}P_2$ scattering wave there is the resonance, and it is not possible to construct $^{3+5}P_2$ potential, which has both resonance and first bound ES using by us Gaussian parametrization of potentials. Therefore, different potentials for discrete and continuous spectra will use at the capture consideration to the first ES. Because, it is known in the MPCM [2,3] that all potentials depend on Young diagrams, so it turns out frequently that potentials of scattering and BS in one partial wave depend on different Young diagrams, since they can be different. Such potential dependence, apparently, was considered in work [20] for the first time. Given above classification according to Young diagrams has only qualitative character, since we have no product tables of Young diagrams, as it was for more light nuclei with $A \leq 8$ [18]. Therefore, in more accurate classification, it is possible that similar situation exists in this system too. Earlier, the same situation took place only in one case – the neutron capture on $^{16}$O [3,21]. Although, other causes that lead to the difference of scattering and BS potentials in this partial wave certainly possible.

Besides excited states there are few resonance states (RS), that is, states at positive energies relatively to the threshold of the $n^{11}$B channel (some states are marked by *italic*, which exist in spectra, but not considered by different causes).

1. First resonance state (1$^{st}$ RS) of $^{12}$B in the $n^{11}$B channel is at the excitation energy of 3.3891(16) MeV or at neutron energy $E_n$ = 20.8(5) keV, it has the width of 1.4 keV in a center of mass (c.m.) and momentum of $J^\pi = 3^-$ (see Table 12.10 in [4]) – it can be matched to the $^{3+5}D_3$ scattering wave with forbidden bound state. We will consider this resonance, because the $E$1 transition the FES $^{3+5}P_2$ of the form $^{3+5}D_3 \rightarrow {}^{3+5}P_2$ is possible for it.

2. Second resonance state (2$^{nd}$ RS) lies at an energy of $E_n$ = 430(10) keV, its width in c.m. equals 37(5) keV and momentum of $J^\pi = 2^+$ [4]. Therefore, it can be matched to the $^{3+5}P_2$ scattering wave with forbidden bound state and the $M$1 transition to the FES $^{3+5}P_2$ is possible for it.

3. Third resonance state (3$^{rd}$ RS) lies at an energy of $E_n$ = 1027(11) keV, its width in c.m. equals 9(4) keV and momentum of $J^\pi = 1^-$ [4] – it can be matched to the $^3S_1$ without FS or $^{3+5}D_1$ scattering wave with forbidden bound state and the $E$1 transition to the FES $^{3+5}P_2$ is possible for it.

*4. Fourth resonance state (4$^{th}$ RS) lies at an energy of $E_n$ = 1.19 MeV, with unknown, but very large width and momentum of $J^\pi = 2^-$ (see Table 12.10 [4]) – it can be matched to the $^5S_2$ or $^{3+5}D_2$ scattering wave and the E1 transition to the FES $^{3+5}P_2$ is possible for it. However, we will not consider such transitions due to unknown width.*

*5. Fifth resonance (5$^{th}$ RS) has an energy of $E_n$ = 1.280(20), with the width of 130(20) keV in c.m. and momentum of 4$^-$ [4]. We will not consider this transition because E1 or M1 transitions to the FES $^{3+5}P_2$ are impossible.*

6. Sixth resonance state (6$^{th}$ RS) is at energy of $E_n$ = 1.780(20) MeV, with the width of 60(20) keV in c.m. and momentum of $J^\pi = 1^+$ can be matched to the $^{3+5}P_1$ *scattering wave with the bound FS (see Table 12.10 [4]) and the M1 transition to the FES $^{3+5}P_2$ is possible for it. However, this state is possible to describe only in the $^5F_1$ wave with FS [1], therefore only E2 transition to the FES is possible and we will not consider it.*

7. Seventh resonance state (7$^{th}$ RS) is at energy of $E_n$ = 2.450(20) MeV, with the width of 110(40) keV in c.m. and momentum of $J^\pi$ = 3$^+$ can be matched to the $^5P_3$ or $^{3+5}F_3$ scattering wave (see Table 12.10 [4]). This state is possible to describe only for the $^{3+5}F_3$ wave [1], and here only E2 transition is possible and we will not consider it.

8. Eighth resonance state (8$^{th}$ RS) is at energy of $E_n$ = 2.580(20) MeV, with the width of 55(20) keV in c.m. and momentum of $J^\pi$ = 3$^-$ can be matched to the $^{3+5}D_3$ scattering wave (see Table 12.10 [4]) and we will consider it because the E1 transition to the FES $^{3+5}P_2$ is possible here.

9. Ninth resonance state (9$^{th}$ RS) lies at an energy of $E_n$ = 2.9 MeV, with unknown width and momentum of $J^\pi$ = 1$^-$ can be matched to the $^3S_1$ scattering wave without FS (see Table 12.10 [4]) and leads to the E1 transition to the FES $^{3+5}P_2$. However, we will not consider such transitions due to unknown width.

10. Tenth resonance state (10$^{th}$ RS) is at energy of $E_n$ = 3.5 MeV, with the width of 140 keV in c.m. and momentum of $J^\pi$ = 1$^+$ can be matched to the $^{3+5}P_1$ scattering wave with the bound FS (see Table 12.10 [4]) and it can lead to the M1 transition to the FES $^{3+5}P_2$. However, in this case only the resonance for the $^5F_1$ wave [1] can be described, and here only E2 transition is possible and we will not consider it.

11. Furthermore, the 11$^{th}$ RS will be the next at energy of $E_n$ = 4.03 MeV, with the unknown width and momentum of $J^\pi$ = 1$^+$. It can be matched to the $^3S_1$ wave without FS and the E1 transitions to the FES $^{3+5}P_2$ are possible for it. But we will not consider it due to the unknown width.

12. The next resonance 12$^{th}$ RS is at energy of $E_n$ = 4.55 MeV, with the width lower than 14 keV in c.m. and unknown momentum, therefore we will not consider it.

13. Furthermore, there is the resonance state (13$^{th}$ RS) with neutron energy of 4.70 MeV, with the width of 45 keV in c.m. and momentum of $J^\pi$ = 2$^-$. It can be matched to the $^5S_2$ wave without FS and the E1 transitions to the FES $^{3+5}P_2$ are possible for it. In this case the $^{3+5}D_2$ wave with FS is possible, which also allows E1 transition to the FES $^{3+5}P_2$.

14. The resonance state (14$^{th}$ RS) at the energy of 4.80 MeV, with the width of 90 keV in c.m. has momentum of $J^\pi$ = 1$^-$. It can be matched to the $^3S_1$ wave without FS and the E1 transitions to the FES $^{3+5}P_2$ are possible for it. In this case the $^{3+5}D_1$ wave with FS is possible, which also allows E1 transition to the FES $^{3+5}P_2$.

15. Higher states are studied till not so closely [4], and we will not consider them. Consequently, it is possible to consider influence of six resonances at energies up to $E_n$ = 5 MeV – these are resonances Nos. 1,2,3,8,13,14. Rests of resonances have either unknown width or momentum, therefore it is impossible to construct unambiguous potentials, as it was done for other resonances [1]. Besides, resonances No. 6, No. 7 and No. 10, as it was said before, are possible to be described only in assumption of the F waves [1], which allow only E2 transitions to the FES and will not be considered.

Spectrum of the described resonance states is listed in Fig. 1b [4].

| $E_n$ (MeV ± keV) | $\Gamma_{cm}$ (keV) | $^{12}B^*$ (MeV) | $l$ | $J^\pi$ |
|---|---|---|---|---|
| 0.0208 ± 0.5 | ≪ 1.4 | 3.3889 | 2 | $3^-$ |
| 0.43 ± 10 | 37 ± 5 | 3.764 | 1 | $2^+$ |
| 1.027 ± 11 | 9 ± 4 | 4.311 | 0 | $1^-$ |
| 1.19 | broad | 4.46 | 0, 2 | $2^-$ |
| 1.28 ± 20 | 130 ± 20 | 4.54 | 2 | $4^-$ |
| 1.78 ± 20 | 60 ± 20 | 5.00 | 1 | $1^+$ |
| 2.45 ± 20 | 110 ± 40 | 5.62 | 1 | $3^+$ |
| 2.58 ± 20 | 55 ± 20 | 5.73 | 2 | $3^-$ |
| 2.9 | broad | 6.0 | 0, 2 | $1^-$ |
| 3.5 | 140 | 6.6 | 1 | $1^+$ |
| 4.03 | broad | 7.06 | 0, 2 | $1^-$ |
| 4.55 | ≤ 14 | 7.54 | > 3 | |
| 4.70 | 45 | 7.68 | 0, 2 | $2^-$ |
| 4.80 | 90 | 7.77 | 0, 2 | $1^-$ |

Fig. 1b. Spectrum of resonance states of $^{12}B$ from Table 12.10 of [4] at energy less than 5 MeV.

As it was shown above, resonance states 3 and 14 can coincide by their moments with the third ES and 13 with the second ES. But it impossible to construct $S$ potentials, which have bound AS, coinciding with one of the ES and having the resonance at the observed excitation energy. Therefore, we will construct these resonance potentials so that they correspond to the $D$ waves with FS and have the resonance at the necessary energy with necessary width, and potentials of the bound excited states correspond to $S$ waves. Furthermore, as a result of these ES and RS, the transition listed in Table 1 will be considered.

Table 1. The list of possible transitions from the $\{^{(2S+1)}L_J\}_i$ state to the different WF components $\{^{(2S+1)}L_J\}_f$ FES $^{3+5}P_2$ of $^{12}B$ at the neutron capture on $^{11}B$ and Gaussian parameters for initial scattering states [1].

| No. | $\{^{(2S+1)}L_J\}_i$ | Transition | $\{^{(2S+1)}L_J\}_f$ | $P^2$ | $V_0$, MeV | $\alpha$, fm$^{-2}$ | $E_r$, keV | $\Gamma_r$, keV |
|---|---|---|---|---|---|---|---|---|
| 1 | $^3S_1$ nonresonance scattering wave. | E1 | $^3P_2$ | 5 | 5.61427 | 0.04 | --- | --- |
| 2 | $^5S_2$ nonresonance scattering wave. | E1 | $^5P_2$ | 5 | 6.70125 | 0.03 | --- | --- |
| 3 | $^3D_1$ resonance at 1.027 MeV – No.3. | E1 | $^3P_2$ | 1/10 | 1611.95103 | 1.25 | 1027 [1027(11)] | 9 [9(4)] |
| | $^5D_1$ resonance at 1.027 MeV – No.3. | | $^5P_2$ | 9/10 | | | | |
| 4 | $^3D_1$ resonance at 4.8 MeV – No.14. | E1 | $^3P_2$ | 1/10 | 4502.245 | 3.5 | 4800 [4800] | 86 [90] |
| | $^5D_1$ resonance at 4.8 MeV – No.14. | | $^5P_2$ | 9/10 | | | | |
| 5 | $^3D_2$ resonance at 4.7 MeV – No.13. | E1 | $^3P_2$ | 3/2 | 6444.382 | 5.0 | 4700 [4700] | 49 [45] |
| | $^5D_2$ resonance at 4.7 MeV – No.13. | | $^5P_2$ | 7/2 | | | | |

| 6 | $^3D_3$ resonance at 2.58 MeV – No.8. | E1 | $^3P_2$ | 42/5 | 654.0477 | 1.1 | 2580 [2580(20)] | 55 [55(20)] |
|---|---|---|---|---|---|---|---|---|
|   | $^5D_3$ resonance at 2.58 MeV – No.8. |   | $^5P_2$ | 28/5 |   |   |   |   |
| 7 | $^3D_3$ resonance at 20.8 keV – No.1. | E1 | $^3P_2$ | 42/5 | 16.0578 (7.96541) | 0.0125 (0.00625) | 20.8 [20.8(5)] | 0.5 [<1.4] (1.3) |
|   | $^5D_3$ resonance at 20.8 keV – No.1. |   | $^5P_2$ | 28/5 |   |   |   |   |
| 8 | $^3P_1$ nonresonance scattering wave. | M1 | $^3P_2$ | 5/2 | 194.68751 | 0.22 | --- | --- |
|   | $^5P_1$ nonresonance scattering wave. |   | $^5P_2$ | 9/2 |   |   |   |   |
| 9 | $^3P_2$ resonance at 430 keV – No.2. | M1 | $^3P_2$ | 15/2 | 11806.017 | 15.0 | 430 [430(10)] | 37 [37(5)] |
|   | $^5P_2$ resonance at 430 keV – No.2. |   | $^5P_2$ | 125/6 |   |   |   |   |

The $P^2$ value determines coefficient in capture cross sections [1]. The number of resonance from the list given above, its energy and width with the given potentials is given for the initial state. The experimental values of energies and width of these resonances [4] are given in square brackets. The alternative parameter values and resonance characteristics with them are given in curly brackets.

Cross sections of some E1 transitions, for example No. 3, No. 4, No. 5, No. 6 and No. 7 from Table 1, can be written as

$$\sigma(E1) = \{\sigma(^3D_1 \to ^3P_2) + \sigma(^5D_1 \to ^5P_2)\}/2 + \{\sigma(^3D_2 \to ^3P_2) + \sigma(^5D_2 \to ^5P_2)\}/2 + \{\sigma(^3D_3 \to ^3P_2) + \sigma(^5D_3 \to ^5P_2)\}/2.$$

The averaging of cross section for the transition from mixed $D_1$, $D_2$ and $D_3$ scattering states to the mixed FES. On the basis of the observed values (it is energy of state and its width [2,3] for resonances) it is possible to construct potential of the $^{3+5}D$ waves, also, as the $^{3+5}P_2$ potential for the FES. Therefore, for example, transitions No. 3 from $^3D_1$ and $^5D_1$ states differ only by spin coefficients in the expression for the cross section (2), given in work [1], and matrix elements are calculated between the same WF, mixed by the channel spin – more closely this problem is given in [2,3] or [22].

The cross section for the M1 transition No. 8 and No. 9 from Table 1 are also written in the form of averaging

$$\sigma(M1) = \{\sigma(^3P_1 \to ^3P_2) + \sigma(^5P_1 \to ^5P_2)\}/2 + \{\sigma(^3P_2 \to ^3P_2) + \sigma(^5P_2 \to ^5P_2)\}/2,$$

because in the initial channel the mixed $^{3+5}P$ waves are used.

## 4. $n^{11}B$ interaction potentials

For all partial potentials, that is, interactions for each orbital momentum L at the given J and S, the Gaussian form was used

$$V(^{2S+1}L_J,r) = -V_0(^{2S+1}L_J)exp\{-\alpha(^{2S+1}L_J)r^2\},$$

where depth $V_0$ and width $\alpha$ of the potential depend on momenta $^{2S+1}L_J$ of each partial wave. In some cases this potential can directly depend from Young diagrams $\{f\}$ and be different in discrete and continuous spectrum, if in such states these diagrams different [19,20].

Furthermore, for constructing FES potential $^{3+5}P_2$ let us determine its AC. For ANC of the form (1) in [23] were obtained value of 0.62(3) fm$^{-1/2}$ and FES spectroscopic factor $S_f = 0.36$ is given. Consequently, the range of dimensioned AC is equal to 0.98–1.08 fm$^{-1/2}$ that for dimensionless AC $C_w$ of the form (3) at $\sqrt{2k_0} = 0.81$ gives an interval of 1.21–1.33 at the average value of 1.27(6). Earlier results from the previous works, specifically, $S_f = 0.55$ and $A_{NC} = 1.34(12)$ from which we obtain $C_w = 2.23(20)$, are given in work [23]. As seen, there is a large dispersion of different results, but we will use new data.

Let us construct FES potential with FS for the first AC

$$V_0 = 213.830142 \text{ MeV and } \alpha = 0.25 \text{ fm}^{-2}. \qquad (4)$$

This potential gives negative AC of -1.24(1), because contains FS stable at the range of 5–17 fm, charge radius of 2.43 fm and mass radius of 2.54 fm at binding energy of -2.416900 MeV with an accuracy of $10^{-6}$ MeV [2,3], which is absolutely coincides with the experimental value [4]. Here and further, the AC error is determined by its averaging over the specified distance interval, and the scattering phase shift of such potential is shown at the bottom of Fig. 2a by the dashed curve.

The values $S_f = 0.33$ and $A_{NC} = 1.28(6)$ fm$^{-1/2}$ are given in [23] for the second ES $^5S_2$, then we have $C_w = 3.0(1)$ at $\sqrt{2k_0} = 0.742$. In such case, it is possible to obtain values No. 2 in Table 1 for parameters of this potential. It leads to the binding energy of -1.6964 MeV, gives AC equals 3.1(1), mass radius of 2.78 fm and charge radius of 2.45 fm, its phase shift is shown in Fig. 2a by the solid curve.

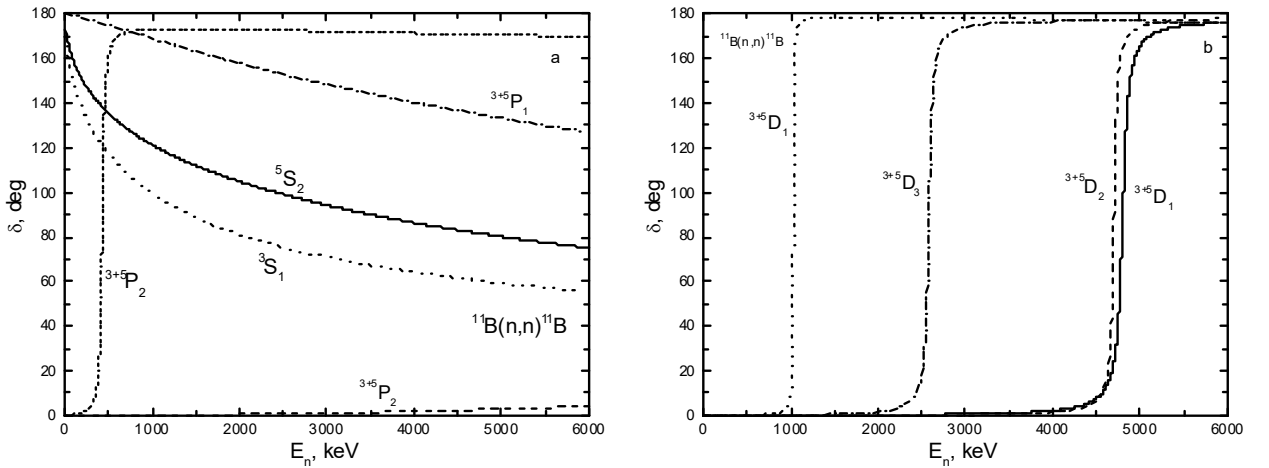

Fig. 2. Phase shifts of the $n^{11}$B elastic scattering at low energies, obtained with potentials, parameters of which are given in the text.

The values $S_f = 0.63$ and $A_{NC} = 1.05(5)$ fm$^{-1/2}$ are given for the third excited $^3S_1$ state in [23], then we have $C_w = 2.2(1)$ at $\sqrt{2k_0} = 0.605$. If to consider further that for the AS for diagram {444} in the $^3S_1$ wave can be bound and corresponds to the third excited state of $^{12}$B at energy -0.7492 MeV relatively to channel threshold, then it is possible to obtain values given as No. 1 in Table 1 for parameters of the $^3S_1$ potential without FS. The binding energy of -0.7492 MeV at ε = $10^{-4}$, absolutely matching with experimental value [4], the charge radius of 2.47 fm, the mass radius of 2.94 fm and dimensionless AC equals 1.9(1) at the range 10–30 fm were obtained with this potential. Shape of the $^3S_1$ scattering phase shift with this potential is shown in the Fig. 2a by the dotted curve. These two $S$ potentials of the ES were used not only for the description of the ESs themselves, but also for description of the $n^{11}$B scattering in these partial waves.

We will use the potential of the GS of $^{12}$B in the $n^{11}$B channel [1] for nonresonance $^{3+5}P_1$ scattering wave. Let us note for the construction of the potential that in works [24–26] the values of 1.13 to 1.35 $fm^{-1}$ were obtained for the ANC of form (1), and in works [4,7,27] the GS spectroscopic factor of 0.69 to 1.30 was given. Consequently, for the range of dimensioned AC we will obtain 0.87–1.96 $fm^{-1}$ or 0.93–1.40 $fm^{-1/2}$, which for dimensionless AC $C_w$ of form (3) at $\sqrt{2k_0}$ = 0.880 gives an interval of 1.06–1.59 at the average value of 1.32(27). More modern data are given in [23], that is $S_f$ = 0.69 and $A_{NC}$ = 1.15(6) $fm^{-1/2}$. Therefore, for the interval of dimensioned AC we will obtain 1.31–1.38 $fm^{-1/2}$, which for dimensionless AC $C_w$ of form (3) gives an interval of 1.49–1.57 at the average value of 1.53(4).

Let us construct the GS potential with FS, which corresponds to the AC from this diapason – its parameters are listed under No. 8 in Table 1. This potential gives negative AC of -1.64(1), because it contains FS, stable in the interval of 5–17 fm, the charge radius of 2.43 fm and the mass radius of 2.53 fm at the binding energy of -3.37000 MeV with an accuracy of $10^{-5}$ MeV [2,3], which is completely coincides with experimental value [4]. Scattering phase shift of this potential is shown in Fig. 2a by the dotted-dashed curve. For comparison of radii it is possible to use the matter radius of 2.41(3) fm, obtained in [28].

It should be noted that in [23] the negative sign of AC is given for both $S$ waves – the value of -1.05(5) $fm^{-1/2}$ for the 3$^{rd}$ ES and -1.28(6) $fm^{-1/2}$ for the 2$^{nd}$ ES. The negative sign of the AC means the presence of node in the WF of relative motion of clusters and the form of these WFs is given in [23]. But, according our classification, there are no bound forbidden states in $S$ waves. According the given above state classification by Young diagrams only $P$ waves have the bound FS. Therefore, for the GS and FES with potential No. 8 and (4) AC has negative sign [1], and for potentials No. 1 and No. 2 – positive.

Parameters No. 9 in Table 1 with one bound FS were obtained for potential of the resonance $^{3+5}P_2$ wave. This potential leads to the resonance energy of $E_n$ = 430(1) keV with the width of 37(1) keV (c.m.), which absolutely coincide with experimental data [4]. The scattering phase shift is equal to 90(1)° for resonance energy. The expression Γ = $2(dδ/dE)^{-1}$ is used for calculation the level width according to the scattering phase shift δ. The shape of the mixed $^{3+5}P_2$ scattering phase shift is shown in Fig. 2a by the frequent dashed curve. Note that accurate values of particle masses and given above value of the constant $\hbar^2/m_0$ were used for calculation of the scattering phase shifts. These values influence very much to location of the considered resonance or BS binding energy of $^{12}$B in the $n^{11}$B channel.

Besides, let us note [1] that if the potential contains $N + M$ forbidden and allowed states, it is subject to generalized Levinson theorem and its phase shift at zero energy starts from π·($N + M$) [19]. However, in Fig. 2a the $P_1$ phase shift having bound FS and bound AS starting from 180°, but not from 360°, and the $P_2$ phase shift starts from 0°, but not from 180° for more traditional representation of results and location of all phase shifts at one figure.

The potentials with FS Nos. 5 and 4 were obtained for $^{3+5}D_2$, $^{3+5}D_1$ resonances at 4.7 and 4.8 MeV correspondingly. The first of them gives width of 49(1) keV, and the second 86(1) keV, which in a good agreement with data [4], and energies accurately coincides with the given above. Phase shifts of these potentials are shown in Fig. 2b by the dashed and solid curves and in resonance equal 90(1)°.

The resonance at 1.027 MeV also can be reproduced only under assumption of the $^{3+5}D_1$ wave and parameters with FS No. 3, which lead to the width of 9(1) keV at the accurately coincident energy and scattering phase shifts shown in Fig. 2b by the dotted curve, which is equal to 90(1)° at resonance.

The resonance at 2.58 MeV in the $^{3+5}D_3$ wave leads to the potential shown in Table 1 as No. 6 at the accurately coincident energy and width 55(1) keV. Phase shift of such potential is shown in Fig. 2b by the dashed-dotted curve.

The potential parameters are listed in Table 1 as No. 7 for the first resonance in the $^{3+5}D_3$ wave at 20.8 keV. They lead to the accurate resonance energy and to the width of 0.5(1) keV in c.m. Let us note that accurate width of this resonance is unknown [4]. Other parameters are given as an alternative option in curly brackets under No. 7, which lead to the accurate resonance energy and to the width of 1.3(1) keV.

As was said earlier, the resonances of positive parity Nos. 6, 7, 10 from the list of resonances at energies 1.78, 2.45 and 3.5 MeV is possible to reproduce only in assumption of the $F$ waves [1]. But in this case, only $E2$ transitions to the FES are possible.

## 5. Total cross section of the neutron capture on $^{11}$B and reaction rate

Calculation results of total cross sections for all transitions from Table 1 with given their potentials of the initial channel, using for the FES potential (4), are shown in Fig. 3a by the black solid curve. Cross sections are calculated at neutron energy of 10 meV to 7 MeV, but as it was said before, the resonances higher 5 MeV were not considered. The green dotted-dashed curve in Fig. 3 shows $E1$ transitions under No. 1 and No. 2 from Table 1. The red dashed curve in Fig. 3 shows calculation result of the $M1$ capture cross section from the resonance $^{3+5}P_2$ wave under No. 9 and nonresonance $^{3+5}P_1$ wave under No. 8 from Table 1. The blue dotted curve shows $E1$ cross section taken into account resonances No. 3 at 1027 keV, No. 4 and No. 5 with resonances at 4.8 and 4.7 MeV that merge at Fig. 3, No. 6 with resonance at 2.58 MeV and No. 7 with resonance at 20.8 keV – the value of the last achieves almost 21mb. The energy range 1 keV – 7 MeV is shown in detail in Fig. 3b and additional experimental data from works [30,31] are given.

Furthermore, the same results for summed cross sections are shown for comparison in Fig. 3c by the black solid curve. Red solid curve in Fig. 3c shows results for the FES (4) and the second option of potential No. 7 from Table 1 with the width of 1.3 keV and cross section maximum of 15 mb at 20.8 keV. The cross section in the maximum decreases by factor of 2, but at other energies notably increases that leads to rise of the reaction rate. Blue solid curve in Fig. 3c shows results for FES (4) and resonance potential No. 7, but with parameters $V_0 = 32.2375215$ MeV and $\alpha = 0.025$ fm$^{-2}$, which leads to the same resonance energy and has the width of 0.2 keV. In the maximum the cross sections are equal almost to 8 mb, and the cross sections themselves have smaller value at all energies. It is seen from here that the first potential No. 7 leads to the highest cross section value at the resonance energy of 20.8 keV and quite satisfies to the experimental requirement for Γ < 1.4 keV.

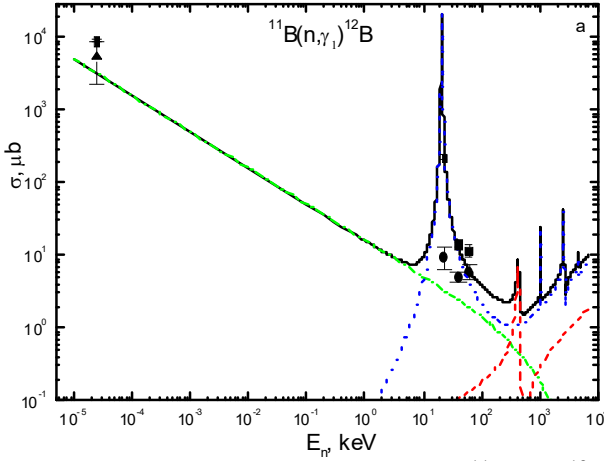

Fig. 3a. Total cross section of the $^{11}$B(n,γ)$^{12}$B radiative capture for transition to the FES in the energy range of $10^{-5}$ keV to 7 MeV. Experimental data: black triangles (▲) – data from work [8] at thermal energy, black square (■) – data from [9] at thermal energy, points (●) – total capture cross sections to the GS from [29], rhombs (♦) – total capture cross sections to the second ES from [29], open triangles (Δ) – total capture cross sections from [31]. Curves – calculation results for different transitions from Table 1 with the given their potentials of the initial channel.

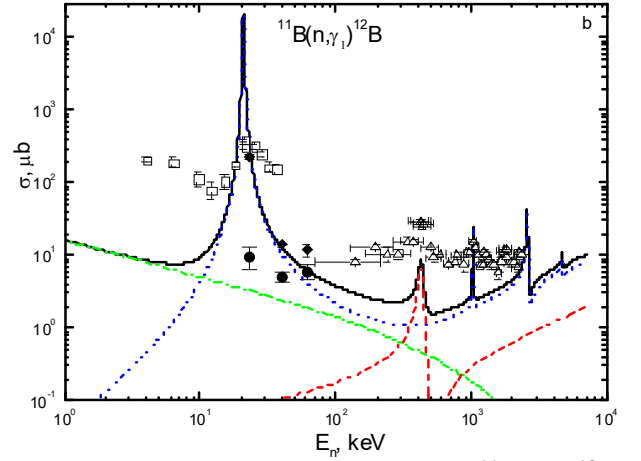

Fig. 3b. Total cross section of the $^{11}$B(n,γ)$^{12}$B radiative capture for transition to the FES in the energy range of 1 keV to 7 MeV. Notations of experimental data as in Fig. 3a.

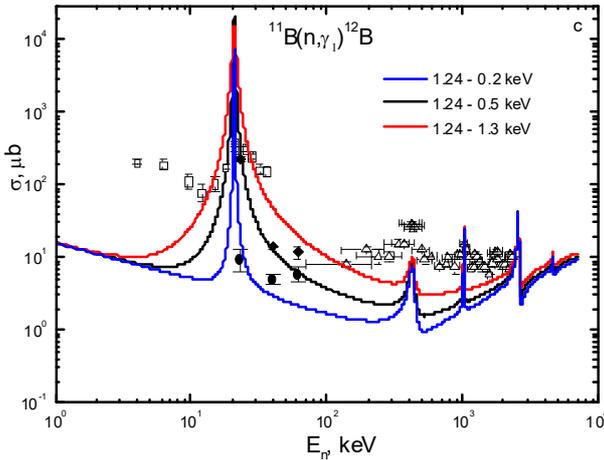

Fig. 3c. Total cross section of the $^{11}$B(n,γ)$^{12}$B radiative capture for transition to the FES in the energy range of 1 keV to 7 MeV. Notations of experimental data as in Fig. 3a. Curves – calculation results for the FES potential (4) with different potentials of the $^{3+5}D_3$ resonance at 20.8 keV.

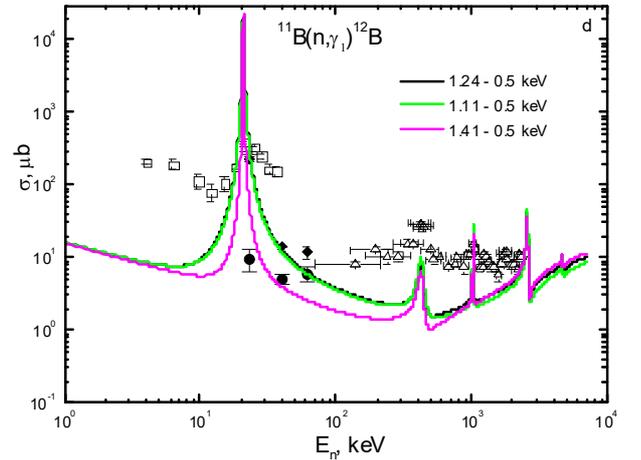

Fig. 3d. Total cross section of the $^{11}$B(n,γ)$^{12}$B radiative capture for transition to the FES in the energy range of 1 keV to 7 MeV. Notations of experimental data as in Fig. 3a. Curves – calculation results for three FES potentials and the first potential of the $^{3+5}D_3$ resonance at 20.8 keV.

The dependence of total cross sections for the first potential of the resonance state

No. 7 from different FES potentials is shown in Fig. 3d. Black curve coincides with results at other Fig. 3 for the FES potential (4). Green curve shows results for the FES potential with parameters $V_0 = 253.829983$ MeV and $\alpha = 0.3$ fm$^{-2}$, which leads to the AC of 1.11, the mass radius of 2.52 fm, the charge radius of 2.42 fm and to the binding energy -2.4169 MeV. Violet curve shows results for the FES potential with parameters $V_0 = 173.68781$ MeV and $\alpha = 0.2$ fm$^{-2}$, which leads to the AC of 1.41, the mass radius of 2.57 fm, the charge radius of 2.43 fm and to the binding energy -2.4169 MeV. It is seen from this figure that results for the FES potential with AC = 1.11 practically coincide with results for potential (4) which AC is equal to 1.24.

As seen from Fig. 3a the calculated cross section practically is a straight line at energies of 10 meV to 10 keV and it can be approximated by a simple function of the next form

$$\sigma_{ap}(\mu b) = \frac{A}{\sqrt{E(keV)}}.$$

The value of constant $A = 16.124$ μb·keV$^{1/2}$ is determined according to one point in calculated cross sections (black solid curve in Fig. 3a) at minimal energy equals 10 meV. The cross section value at thermal energy 25.3 meV is equal to 3.2 mb. The module

$$M(E) = \left| [\sigma_{ap}(E) - \sigma_{theor}(E)] / \sigma_{theor}(E) \right|$$

of relative deviation of the calculated theoretical cross section ($\sigma_{theor}$) and approximation ($\sigma_{ap}$) of this cross section by the given above function in the range 10 keV does not exceed 0.1%. Earlier in our work [1] the cross sections in the range of 7.3 to 8.4 mb were obtained cross sections for the capture to the GS at thermal energy for different types of potentials. Therefore, for the thermal cross section taken into account the capture to the GS and the FES it is possible to obtain values of 10.5–11.6 mb at new experimental value 9.09(10) mb [9].

Furthermore, the reaction rate of the neutron capture on $^{11}$B was calculated, which in terms of cm$^3$mol$^{-1}$s$^{-1}$ can be represented in form [10]

$$N_A \langle \sigma v \rangle = 3.7313 \cdot 10^4 \mu^{-1/2} T_9^{-3/2} \int_0^\infty \sigma(E) E \exp(-11.605 E / T_9) dE,$$

where $E$ is in MeV, total cross section $\sigma(E)$ is measured in μb, μ is the reduced mass in amu, and $T_9$ is the temperature in $10^9$ K [10]. The calculated total cross section is used for determination such integral for 7000 points in the energy range of $10^{-5}$ to $7 \cdot 10^3$ keV.

Results for reaction rate are shown in Fig. 4 by the black solid curve. This curve corresponds to the same curve in Fig. 3 for the FES potential (4) and for the first option of potential No. 7 from Table 1. Red dotted curve shows rate of this reaction without accounting the first resonance. As seen from Fig. 4 the resonance accounting at energy up to 5 MeV, especially the first resonance at 20.8 keV, greatly influence to the total cross sections in this range and highly work on the shape of reaction rate.

For comparison the blue dotted-dotted-dashed curve shows the results from work [7]. The obtained by us reaction rate of the neutron capture on $^{11}$B to the GS of $^{12}$B [1] is shown by the green dashed curve. Parametrization results for this reaction rate are given in our previous work [1]. Red solid curve shows the results for the second option of potential No. 7, shown in Table 1 by curly brackets – they correspond to the same curve in Fig. 3c. Blue solid curve shows the results for potential No. 7 at the width of 0.2 keV – they correspond to the same curve in Fig. 3c.

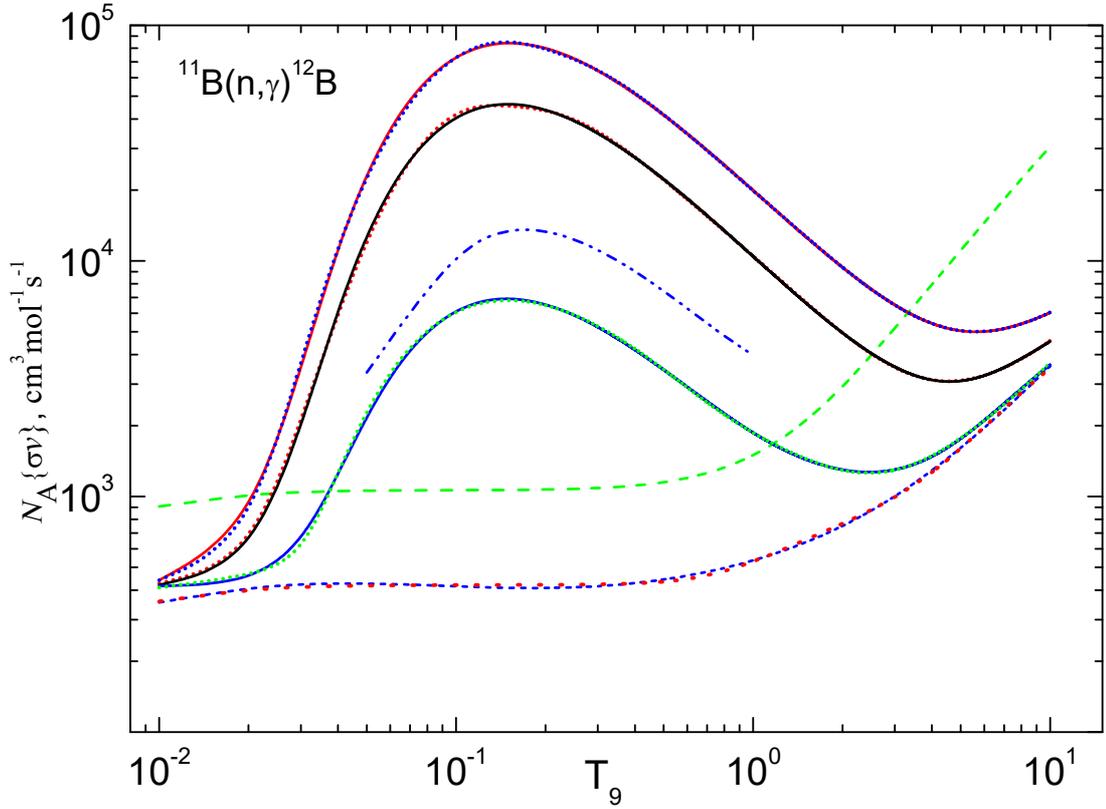

Fig. 4. Reaction rate of the radiative neutron capture on $^{11}$B to the FES. Notations of the curves are given in the text.

Furthermore, the parametrization of the reaction rate was carried out by the red dotted curve in Fig. 4 using expression of the form [32]

$$N_A \langle \sigma v \rangle = a_1 / T_9^{2/3} \exp(-a_2 / T_9^{1/3})(1.0 + a_3 T_9^{1/3} + a_4 T_9^{2/3} + a_5 T_9 + a_6 T_9^{4/3} + a_7 T_9^{5/3})$$

with parameters listed in Table 2.

Table 2. Parameters of the analytical parametrization of the reaction rate

| No. | 1 | 2 | 3 | 4 | 5 | 6 | 7 |
|---|---|---|---|---|---|---|---|
| $a_i$ | 5.80196 | 0.51371 | 241.0119 | -680.9735 | 1105.41 | -763.5066 | 251.3789 |

$\chi^2$ value is equal to 0.08 at 5% errors of the calculated reaction rate, and results are shown in Fig. 4 by the blue frequent dashed curve.

Besides, the parametrization of the reaction rate was carried out by the expression of the form.

$$N_A \langle \sigma v \rangle = a_1 / T_9^{2/3} \exp(-a_2 / T_9^{1/3})(1.0 + a_3 T_9^{1/3} + a_4 T_9^{2/3} + a_5 T_9 + a_6 T_9^{4/3} + a_7 T_9^{5/3} + a_8 T_9^{7/3}) +$$
$$+ a_9 / T_9^{1/2} \exp(-a_{10} T_9^{1/2}) + a_{11} / T_9 \exp(-a_{12} T_9) + a_{13} / T_9^{3/2} \exp(-a_{14} T_9^{3/2}) + a_{15} / T_9^2 \exp(-a_{16} T_9^2) +$$
$$+ a_{17} T_9^{a_{18}}$$

It is shown by the black solid curve in Fig. 4, which takes into account the resonance at 20.8 keV. The parametrization parameters are given in Table 3.

Table 3. Parameters of the analytical parametrization of the reaction rate

| No. | 1 | 2 | 3 | 4 | 5 | 6 | 7 | 8 | 9 |
|---|---|---|---|---|---|---|---|---|---|
| $a_i$ | 38.44148 | 2.45224 | 36104.94 | -19086.46 | 674.7674 | 3417.136 | -1084.624 | 31.39074 | -349835 |
| No. | 10 | 11 | 12 | 13 | 14 | 15 | 16 | 17 | 18 |
| $a_i$ | 1.56469 | 38296.83 | 0.21875 | -22274.84 | 0.16279 | 5834.174 | 0.08573 | 2.79219E-12 | -7.05298 |

Results of such parametrization are shown in Fig. 4 by the red frequent dotted curve, and the $\chi^2$ value is equal to 0.02 at 5% errors of the calculated reaction rate.

The reaction rate shown in Fig. 4 by the red solid curve is parametrized by the same form with parameters listed in Table 4.

Table 4. Parameters of the analytical parametrization of the reaction rate

| No. | 1 | 2 | 3 | 4 | 5 | 6 | 7 | 8 | 9 |
|---|---|---|---|---|---|---|---|---|---|
| $a_i$ | 38.04286 | 2.4494 | 36104.94 | -19086.46 | 732.8953 | 3417.136 | -1083.635 | 30.73185 | -350638.7 |
| No. | 10 | 11 | 12 | 13 | 14 | 15 | 16 | 17 | 18 |
| $a_i$ | 1.56571 | 47661.85 | 0.19211 | -21753.64 | 0.2316 | 5722.362 | 0.13739 | 3.18323E-12 | -7.03645 |

Results of such parametrization are shown in Fig. 4 by the blue frequent dotted curve, and the $\chi^2$ value is equal to 0.01 at 5% errors of the calculated reaction rate.

Some other form

$$N_A \langle \sigma v \rangle = a_1 / T_9^{2/3} \exp(-a_2 / T_9^{1/3})(1.0 + a_3 T_9^{1/3} + a_4 T_9^{2/3} + a_5 T_9 + a_6 T_9^{4/3} + a_7 T_9^{5/3} + a_8 T_9^{7/3}) +$$
$$+ a_9 / T_9^{1/2} \exp(-a_{10} T_9^{1/2}) + a_{11} / T_9 \exp(-a_{12} T_9) + a_{13} / T_9^{3/2} \exp(-a_{14} T_9^{3/2})$$

is used by us for parametrization of the reaction rate, shown in Fig. 4 by the blue solid curve with parameters listed in Table 5.

Table 5. Parameters of the analytical parametrization of the reaction rate

| No. | 1 | 2 | 3 | 4 | 5 | 6 | 7 |
|---|---|---|---|---|---|---|---|
| $a_i$ | 0.6448 | -0.41482 | 959.8919 | -11735.97 | 3365.214 | 3876.845 | -1725.825 |
| No. | 8 | 9 | 10 | 11 | 12 | 13 | 14 |
| $a_i$ | 131.159 | 3402.191 | 0.01741 | 3813.454 | 0.1612 | 2399.866 | 2.14164 |

Results of such parametrization are shown in Fig. 4 by the green frequent dotted curve, and the $\chi^2$ value is equal to 0.03 at 5% errors of the calculated reaction rate.

## 6. Conclusion

The used FES potential of $^{12}$B in the $n^{11}$B channel previously coordinated with its characteristics, including asymptotic constant and binding energy. The interactions of the second and third excitation states of $^{12}$B that bound in the $n^{11}$B channel were used for $S$ scattering potentials. Accounting of resonances above the first and up to 5 MeV increases total cross sections in this energy region and reaction rate at temperatures 0.5–0.7 $T_9$ becomes rise and increasing approximately by factor of 10. Accounting of the first resonance absolutely changes the value of reaction rate, involving it to the resonance shape.

Obtained calculated cross sections and structure of resonances in cross sections can be considered as a prediction of results for total cross sections and as a stimulus for new experimental studies of this reaction. They can essentially influence to the calculation results for efficiency of $^{12}$B in the primordial nucleosynthesis at high temperatures.

In the following our papers, as an addition to the present work [1], the radiative capture is considered to other three ESs of $^{12}$B, which lie lower the threshold of the $n^{11}$B channel and are given above in the list of the ESs.

## Acknowledgements

This work was supported by the Grant of Ministry of Education and Science of the Republic of Kazakhstan through the program BR05236322 "Study reactions of thermonuclear processes in extragalactic and galactic objects and their subsystems" in the frame of theme "Study of thermonuclear processes in stars and primordial nucleosynthesis" through the Fesenkov Astrophysical Institute of the National Center for Space Research and Technology of the Ministry of Defence and Aerospace Industry of the Republic of Kazakhstan (RK).

## References

1. Dubovichenko S B, Burkova N A, Dzhazairov-Kakhramanov A V and Tkachenko A S https://arxiv.org/abs/1907.06832v1
2. Dubovichenko S B 2015 *Thermonuclear processes in Stars and Universe*. Second English edition (Saarbrucken: Scholar's Press.) 332p.; https://www.scholars-press.com/catalog/details/store/gb/book/978-3-639-76478-9/Thermonuclear-processes-in-stars; Dubovichenko S B 2019 *Thermonuclear processes in Stars and Universe*. Fourth Russian edition, corrected and enlarged (Saarbrucken: Lambert Academy Publ. GmbH&Co. KG.) 495p. ISBN 978-620-0-25609-6
3. Dubovichenko S.B. 2019 *Radiative neutron capture. Primordial nucleosynthesis of the Universe*. First English edition (Berlin: Walter de Gruyter GmbH); ISBN 978-3-11-061784-9, https://doi.org/10.1515/9783110619607-201
4. Kelley J H, Purcell J E and Sheu C G 2017 *Nucl. Phys. A* **968**, 71-253
5. Dubovichenko S B, Dzhazairov-Kakhramanov A V, Burkova N A and Tkachenko A S 2017 *Russ. Phys. J.* **60**, 666–677; Dubovichenko S B and Burkova N A 2014 *Mod. Phys. Lett. A* **29**, 1450036(14p.)
6. Ajzenberg-Selove F. 1990 *Nucl. Phys. A* **506**, 1-158
7. Lee H Y et al 2010 *Phys. Rev. C* **81**, 015802(8p.)
8. Mughabghab S F 2006 *Atlas of neutron resonances*, National Nuclear Data Center,

Brookhaven, National Laboratory, Upton, USA, 1008p
9. Firestone R B and Revay Zs 2016 *Phys. Rev. C* **93**, 054306
10. Angulo C *et al* 1999 *Nucl. Phys. A* **656**, 3–183
11. Fundamental Physical Constants 2018; http://physics.nist.gov/cgi-bin/cuu/Value?mud|search_for=atomnuc!
12. Kelley J H *et al* 2012 *Nucl. Phys. A* **880**, 88–233
13. Nucl. Wallet Cards 2015; http://cdfe.sinp.msu.ru/services/ground/NuclChart_release.html
14. Dubovichenko S B and Dzhazairov-Kakhramanov A V 1997 *Phys. Part. Nucl.* **28**, 615–641
15. Dubovichenko S B, Dzhazairov-Kakhramanov A V and Burkova N A 2019 *Int. J. Mod. Phys. E* **28**, 1930004(49p.)
16. Mukhamedzhanov A M and Tribble R E 1999 *Phys. Rev. C* **59**, 3418–3424
17. Plattner G R and Viollier R D 1981 *Nucl. Phys. A* **365**, 8–12
18. Itzykson C and Nauenberg M 1966 *Rev. Mod. Phys.* **38**, 95–120
19. Nemets O F *et al* 1988 *Nucleon Association in Atomic Nuclei and the Nuclear Reactions of Many Nucleons Transfers* (Kiev: Naukova Dumka) 488p. (in Russian).
20. Neudatchin V G *et al* 1992 *Phys. Rev. C* **45**, 1512–1527
21. Dubovichenko S B 2014 *Russ. Phys. J.* **57**, 498–508
22. Dubovichenko S B and Dzhazairov-Kakhramanov A V 2015 *Nucl. Phys. A* **941**, 335–363
23. Belyaeva L T *et al* 2018 *Phys. Rev. C* **98**, 034602(10p.)
24. Lin Cheng-Jian *et al* 2001 *Chin. Phys. Lett.* **18**, 1183–1185
25. Guo B *et al* 2007 J. Phys. G: Nucl. Part. Phys. **34**, 103–114
26. Timofeyuk N K 2013 *Phys. Rev. C* **88**, 044315(11p.)
27. Cook J, Stephens M N and Kemper K W 1987 *Nucl. Phys. A* **466**, 168–188.
28. Estradé A *et al* 2014 *Phys. Rev. Lett.* **113**, 132501(5p.)
29. Igashira M *et al* 1964 Measurements of keV-neutrons capture gamma rays, Conf. Meas. Calc. and Eval. of Photon Prod. Data, Bologna, P.269–280
30. Imnof W L *et al* 1962 *Phys. Rev.* **125**, 1334–1336
31. Mooring F P and Monahan J E 1969 *Phys. Rev.* **178**, 1612–1615
32. Caughlan G R and Fowler W A 1988 *Atom. Data Nucl. Data Tab*. **40**, 283–334